\documentclass[conference]{IEEEtran}
\IEEEoverridecommandlockouts
\usepackage{cite}
\usepackage{amsmath,amssymb,amsfonts}
\usepackage{algorithmic}
\usepackage{graphicx}
\usepackage{textcomp}
\usepackage{xcolor}
\usepackage{etoolbox}
\usepackage{multirow}
\usepackage{makecell}
\usepackage{float}
\usepackage{subcaption}

\patchcmd{\quote}{\rightmargin}{\leftmargin 1.5em \rightmargin}{}{}

\def\BibTeX{{\rm B\kern-.05em{\sc i\kern-.025em b}\kern-.08em
    T\kern-.1667em\lower.7ex\hbox{E}\kern-.125emX}}
\begin{document}

\title{Action Word Prediction for Neural Source\\Code Summarization}

		
\author{\IEEEauthorblockN{Sakib Haque\IEEEauthorrefmark{1},
		Aakash Bansal\IEEEauthorrefmark{1}, Lingfei Wu\IEEEauthorrefmark{2} and
		Collin McMillan\IEEEauthorrefmark{1}\\
		\IEEEauthorblockA{\IEEEauthorrefmark{1}Dept. of Computer Science,
			University of Notre Dame, Notre Dame, IN, USA \\
			\{shaque, abansal1, cmc\}@nd.edu}
		\IEEEauthorblockA{\IEEEauthorrefmark{2}IBM Research,
			Yorktown Heights, NY, USA \\
			\{wuli\}@us.ibm.edu}}}

\maketitle

\begin{abstract}
Source code summarization is the task of creating short, natural language descriptions of source code.  Code summarization is the backbone of much software documentation such as JavaDocs, in which very brief comments such as ``adds the customer object'' help programmers quickly understand a snippet of code.  In recent years, automatic code summarization has become a high value target of research, with approaches based on neural networks making rapid progress.  However, as we will show in this paper, the production of good summaries relies on the production of the action word in those summaries: the meaning of the example above would be completely changed if ``removes'' were substituted for ``adds.''  In this paper, we advocate for a special emphasis on action word prediction as an important stepping stone problem towards better code summarization -- current techniques try to predict the action word along with the whole summary, and yet action word prediction on its own is quite difficult.  We show the value of the problem for code summaries, explore the performance of current baselines, and provide recommendations for future research.
\end{abstract}

\begin{IEEEkeywords}
neural networks, source code summarization, automatic documentation generation, AI in SE
\end{IEEEkeywords}

\section{Introduction}
The task of creating short, natural language descriptions of source code has come to be known as ``source code summarization.''  Code summarization is the backbone of a plethora of documentation such as JavaDocs~\cite{kramer1999api}, in which the natural language description (the ``summary'') provides a quick way for programmers to understand the software's components.  Very often, these summaries are written for subroutines, so that programmers can read that a subroutine e.g. ``computes the dot product of two vectors'' rather than interpret the source code itself.  Traditionally, programmers write these summaries around the time they write the code, to help other programmers in understanding that code.

\emph{Automatic} code summarization has been a dream of software engineering researchers for decades.  Forward~\emph{et al.}~\cite{forward2002relevance} observed almost 20 years ago that ``software professionals value technologies that improve automation of the documentation process,'' and ``that documentation tools should seek to better extract knowledge from core resources.''  Efforts in this direction have begun to bear fruit, especially in the last five years with the introduction of neural models for code summarization.  A confluence of work in both the AI and SE communities has pushed the state-of-the-art to a point where real-world automatic code summarization seems within reach.


Yet, as we will show in this paper, very often these techniques owe their good performance on their ability to predict the \emph{first word} of the summary.  Some of the reasons for this are technical: Existing techniques tend to be based on an encoder-decoder architecture (e.g. seq2seq, graph2seq) in which the output summary is predicted one word at a time.  The first word is predicted first, then that first prediction is used to predict the second word, and so on.  If the first word is wrong, the model can have a hard time recovering.  This situation can be exaggerated by the aggressive use of attention mechanisms (as in Transformer-based models~\cite{ahmad2020transformer}), which can attend previous words in the predicted summary to parts of the source code.  Often each subsequent word depends more and more on the previous predictions.

A more fundamental reason the first word is important is that the first word tends to be the action word in code summaries.  As we will show (and in line with style guides~\cite{kramer1999api, MSStyleGuide}), summaries usually fall into a pattern where the action word not only occurs first, but sets the tone for the whole summary.  Consider examples such as ``initializes the microphone for the web conference'', ``sets the current speaker's volume'', and ``sorts the list of connected users.''  A lot of information is communicated just by knowing that the code initializes, sets, or sorts.  The rest of the summary depends on that information, begging the question: initializes/sets/sorts \emph{what?}

The importance of early predictions in text generation models has been recognized in the NLP community for years, with several proposed technical workarounds e.g. beam search and alternative training strategies.  Meanwhile, the prevalence of verb-direct object patterns in code summaries has long been observed in SE literature~\cite{jiang2017automatically}.  What is not yet recognized is the special importance of the action word in source code summarization, and how to leverage this importance to create better summaries overall.

Clues about how to leverage this importance can be observed from the progression of the literature on neural code summarization.  As we show in Section~\ref{sec:bg}, one strong consensus is that code structure helps ``somehow.''  Almost immediately after the first applications of neural models to code summarization, efforts started focusing on how to combine structural information such as the abstract syntax tree with the text from the code (after all, combining structure and text has a long history in SE literature~\cite{biggerstaff1993concept}).


What we observe, in a nutshell, is that neural models that perform well either 1) extract the action word from the text such as the function name e.g. ``sorts'' for function ``sortSpeakers'', or 2) use the structure to help detect what action word to use.  This second case is possible because there are different types of functions that tend to have similar structure.  A simple example is to compare getters and setters.  Getters tend to have no parameters and return something, while setters tend to have a parameter and no return.  Even without any text at all, it is possible to detect whether a function is a getter or a setter just by looking at the code structure.  Once the first word is selected accurately, the model has a much better chance at writing the rest of the summary.

In this paper, we propose the problem of \textbf{action word prediction} as a stepping stone towards better neural models of code summarization.  We demonstrate several advantages to targeting action word prediction in conjunction with code summarization, namely: 1) good action word prediction leads to good summary prediction overall, 2) prediction of the action word is often possible from the code structure alone, which avoids a dependence on good internal documentation, and 3) action word prediction is 10-20x faster than code summarization, which is important in practice because training times range 10-15 minutes instead of 1-5 hours.

We propose several configurations of action word prediction which we find to be especially useful.  We build several baselines inspired by recent literature, create two large datasets, and demonstrate baseline performance on these datasets.  We make recommendations about good practice for using action word prediction as part of evaluation of code summarization techniques.  We release all datasets and code to the public via our online appendix (see Section~\ref{sec:appendix}).

\section{Problem Summary}
\label{sec:problem}

We call the problem we target in this paper ``action word prediction.''  An action word is a verb that describes what changes one actor performs to another~\cite{hauk2008time, hirsh2010action}.  Typically code summaries have one action word that broadly classifies what the code does (gets, sets, initializes, sorts).  The problem definition is essentially: given a component of source code, predict the action word to be used in the summary.  Action word prediction is necessary to code summarization, and special emphasis on this part of the problem serves as a key stepping stone towards better code summarization.

This work follows a long tradition of creating smaller academic problems that lead to progress in bigger problems.  The intellectual heritage of this tradition was neatly summarized in 1950 in Claude Shannon's famous paper on the academic problem of automatically playing chess~\cite{shannon1950xxii}:

\vspace{0.02cm}
{\small
\begin{quote}
``Although perhaps of no practical importance, the question is of theoretical interest, and it is hoped that a satisfactory solution of this problem will act as a wedge in attacking other problems of a similar nature and of greater significance.''
\end{quote}
}
\vspace{0.01cm}

In other words, even though action word prediction is not necessarily useful to practitioners today, it is still a useful problem to address because of its necessity and value in solving problems that are.
\section{Background and Related Work}
\label{sec:bg}

\begin{figure}[!b]
	{\small
		\vspace{-0.4cm}
		\begin{tabular}{p{3.9cm}p{0.4cm}p{0.4cm}p{0.4cm}p{0.4cm}p{0.4cm}}
			& IR          & M          & T          & A & S                \\
			\textcolor{white}{*}Haiduc~\emph{et al.}~(2010)~\cite{haiduc2010use}					& x	&   &   &  &   \\ 
			\textcolor{white}{*}Sridhara~\emph{et al.}~(2011)~\cite{sridhara2011automatically}		&   & x	& x &  &   \\
			\textcolor{white}{*}Rastkar~\emph{et al.}~(2011)~\cite{rastkar2011generating}			& x & x	& x &  &   \\
			\textcolor{white}{*}DeLucia~\emph{et al.}~(2012)~\cite{de2012using}						& x &  	&   &  &   \\
			\textcolor{white}{*}Panichella~\emph{et al.}~(2012)~\cite{panichella2012mining}			& x & x	&   &  &   \\
			\textcolor{white}{*}Moreno~\emph{et al.}~(2013)~\cite{moreno2013automatic}				& x &  	& x &  &   \\
			\textcolor{white}{*}Rastkar~\emph{et al.}~(2013)~\cite{rastkar2013did}					& x &  	&   &   &  \\
			\textcolor{white}{*}McBurney~\emph{et al.}~(2014)~\cite{mcburney2014automatic}								&   & x	& x &  &   \\
			\textcolor{white}{*}Rodeghero~\emph{et al.}~(2014)~\cite{rodeghero2014improving}							& x &  	&   &  &   \\
			\textcolor{white}{*}Rastkar~\emph{et al.}~(2014)~\cite{rastkar2014automatic}			&   & x	&   &  &   \\
			\textcolor{white}{*}Cort{\'e}s-Coy~\emph{et al.}~(2014)~\cite{cortes2014automatically}	& x &  	&   &  &   \\
			\textcolor{white}{*}Moreno~\emph{et al.}~(2014)~\cite{moreno2014automatic}				& x &  	&   &  &   \\
			\textcolor{white}{*}Oda~\emph{et al.}~(2015)~\cite{oda2015learning}						&   &  	&   & x  & \\
			\textcolor{white}{*}Abid~\emph{et al.}~(2015)~\cite{abid2015using}						&   & x	& x &   &  \\
			\textcolor{white}{*}Iyer~\emph{et al.}~(2016)~\cite{iyer2016summarizing}				&   &  	&   & x  & \\
			\textcolor{white}{*}McBurney~(2016)~\cite{mcburney2016automated}											& x & x	&   &  &   \\
			\textcolor{white}{*}Zhang~\emph{et al.}~(2016)~\cite{zhang2016towards}					&   & x	& x &  &   \\
			\textcolor{white}{*}Rodeghero~\emph{et al.}~(2017)~\cite{rodeghero2017detecting}							&   & x	&   &  &   \\
			\textcolor{white}{*}Fowkes~\emph{et al.}~(2017)~\cite{fowkes2017autofolding}			& x &  	&   &  &   \\
			\textcolor{white}{*}Badihi~\emph{et al.}~(2017)~\cite{badihi2017crowdsummarizer}		&   & x	& x &  &   \\
			\textcolor{white}{*}Loyola~\emph{et al.}~(2017)~\cite{loyola2017neural}					&   &  	&   & x  & \\
			\textcolor{white}{*}Lu~\emph{et al.}~(2017)~\cite{lu2017learning}						&   &  	&   & x  & \\
			\textcolor{white}{*}Jiang~\emph{et al.}~(2017)~\cite{jiang2017automatically}								&   &  	&   & x  & \\
			\textcolor{white}{*}Hu~\emph{et al.}~(2018)~\cite{hu2018summarizing}					&   &  	&   & x  & \\
			\textcolor{white}{*}Hu~\emph{et al.}~(2018)~\cite{hu2018deep}							&   &  	&   & x  & x \\
			\textcolor{white}{*}Allamanis~\emph{et al.}~(2018)~\cite{allamanis2018learning}			&   &  	&   & x  & x \\
			\textcolor{white}{*}Wan~\emph{et al.}~(2018)~\cite{wan2018improving}					&   &  	&   & x  & x \\
			\textcolor{white}{*}Liang~\emph{et al.}~(2018)~\cite{liang2018automatic}				&   &  	&   & x & x \\
			\textcolor{white}{*}Alon~\emph{et al.}~(2019)~\cite{alon2019code2seq, alon2019code2vec}					&   &  	&   & x & x \\
			\textcolor{white}{*}Gao~\emph{et al.}~(2019)~\cite{gao2019neural}						&   &  	&   & x &  \\
			\textcolor{white}{*}LeClair~\emph{et al.}~(2019)~\cite{leclair2019neural}									&   &  	&   & x & x \\
			\textcolor{white}{*}Mesbah~\emph{et al.}~(2019)~\cite{mesbah2019deepdelta}						&   &  	&   & x & x \\
			\textcolor{white}{*}Nie~\emph{et al.}~(2019)~\cite{nie2019framework}						&   &  	&   & x & x \\
			\textcolor{white}{*}Haque~\emph{et al.}~(2020)~\cite{haque2020improved}						&   &  	&   & x & x \\
			\textcolor{white}{*}Haldar~\emph{et al.}~(2020)~\cite{haldar2020multi}						&   &  	&   & x & x \\
			\textcolor{white}{*}Ahmad~\emph{et al.}~(2020)~\cite{ahmad2020transformer}						&   &  	&   & x & x 
		\end{tabular}
	}
	\vspace{0.1cm}
	\caption{\small{Selection of closely-related, peer-reviewed publications in the previous ten years.  Column $IR$ indicates if the approach is based on Information Retrieval.  $M$ indicates manual features/heuristics.  $T$ indicates templated natural language.  $A$ indicates Artificial Intelligence (usually Neural Network) solutions. $S$ means structural data such as the AST is used (for AI-based models).}}
	\label{tab:screlated}
\end{figure}

The related work of this paper includes any work that seeks to generate natural language descriptions from source code.  While an exhaustive survey of this work is beyond the scope of this paper, we provide a snapshot of the previous ten years in Figure~\ref{tab:screlated} below.  This list includes only the first appearance of ideas (i.e. not journal extensions) in peer-reviewed, full-length technical papers, and is narrowly defined as approaches in which the input is source code and the output is natural language (detailed history of code summarization and neural representations of code for other purposes has been chronicled by several surveys~\cite{chen2018best, allamanis2017survey, ml4codewebsite, song2019survey, nazar2016summarizing}).  Yet a trend is immediately clear:  A first generation of techniques relied on Information Retrieval (IR) or manually-defined heuristics to extract words from source code (for example, by using TF/IDF to pick the top-$n$ words~\cite{haiduc2010use, hill2009automatically}), and templates to place those words into readable sentences (exemplars in this category are presented in~\cite{sridhara2011automatically, mcburney2014automatic}).  An important strength in this first generation is that they tended to be built on a solid foundation of empirical evidence of what is important to programmers.  For example, Moreno~\emph{et al.}~\cite{moreno2013automatic} designed sentence templates for Java classes based on specific studies of programmers' needs in documentation.  However, a weakness is that it was infeasible to write manual heuristics capable of considering the huge range of program behavior.  The response was that between 2015 and 2017, a second generation of approaches based on AI (specifically, neural networks) rose to prominence.  These approaches were enabled by complementary research in mining software repositories that provided access to datasets with millions of examples of code and summaries~\cite{Lopes+Bajracharya+Ossher+Baldi:2010, leclair2019recommendations}, as well as evidence that source code is ``natural''~\cite{hindle2012naturalness} in that it follows patterns similar to natural language.

The workhorse of second generation code summarization approaches is the encoder-decoder neural architecture in which the source code is the encoded input and the summary is the decoded output.  A majority use this technology off-the-shelf from the NLP research area for machine translation~\cite{belinkov2019analysis}, albeit with different preprocessing steps intended to highlight different areas of code such as Iyer~\emph{et al.}~\cite{iyer2016summarizing}.  The trend in the past two years has been to squeeze ever more information from the source code, with the Abstract Syntax Tree (AST) being by far the most popular target.  Hu~\emph{et al.}~\cite{hu2018deep} and LeClair~\emph{et al.}~\cite{leclair2019neural} explore different configurations of using the AST, Alon~\emph{et al.}~\cite{alon2018code2seq} use static paths in the tree, and Allamanis~\emph{et al.}~\cite{allamanis2018learning} propose modeling the AST with a graph neural net. 

Hints as to \emph{why} structure helps code summarization were provided by LeClair~\emph{et al.}~\cite{leclair2019neural} in what they called a ``challenge'' experiment.  The experiment was essentially an extreme ablation study in which only the AST was available to the model for training and test -- no text from the source code at all.  The study found performance in the range of 9-10 BLEU points (compared to 19-20 for AST+text configurations), which is actually quite remarkable considering the extreme experimental conditions.  What they found is that the model recognizes structural features to determine that a function e.g. deinitializes data structures, even if no details about what is deinitialized are available (see Example 2 in~\cite{leclair2019neural}).

The same year and working independently, Alon~\emph{et al.}~\cite{alon2019code2seq} found additional evidence that structure helps identify a general purpose behind functions, even if the model must rely on text data to understand specific details.  In the example in Figure 1(b) of~\cite{alon2019code2seq} they demonstrate how a path in the AST helps the model decide that a function replaces some information with some other information.  That example is used as motivation for their path-based AST summarization approach.  While these papers do not say so explicitly, essentially they are showing how structure helps determine the action word of the summary (\emph{closes} a network connection, \emph{replaces} a string in a text file), while remaining dependent on the textual data in a code to write the rest of the summary.

At a high level, the trend in code summarization research is to leverage more domain-specific knowledge, while moving away from models designed for text processing.  Our goal in this paper is to help focus this trend.

\section{Action Word Prediction}
\label{sec:awp}
\vspace{-0.1cm}

In this section, we provide our definition of action word prediction and advocate for a focus on this problem towards better source code summarization.  The crux of our argument is that action words are important parts of code summaries, so code summarization tools need to predict action words anyway.  Since action words are fundamental to the meaning of code summaries, it is especially important to predict them well.  We also observe that action words have an outsize effect on current AI-based models for code summarization, so future advances are likely to require improved action word prediction.  We build our case in this section based on analysis of state-of-the-art techniques over a curated dataset of over 2.1m Java method summaries~\cite{leclair2019recommendations} as well as 1.1m C/C++ function summaries that we collected from GitHub for this paper.  We use four recent approaches (\cite{iyer2016summarizing, leclair2019neural, xu2018graph2seq, alon2018code2seq}, described in more detail in Section~\ref{sec:baselines}) and BLEU scores of those approaches over these datasets, to be consistent with evaluation practice in those papers. 

\textbf{Definitions} An ``action word'' in a source code summary is a word that broadly classifies what the code does (as mentioned in Section~\ref{sec:problem}).  This almost always means a verb.  Common programming parlance separates code functions into e.g. ``getters'', ``setters'', ``initializers''.  As we will show, this parlance is fundamental to how programmers write code summaries: programmers tend to write summaries that contain only one action word, that word tends to occur in the first position of the sentence, and very often it is the only verb in the sentence.

\begin{figure}[!b]
	\centering
	{\small

    \vspace{-0.3cm}
	\begin{tabular}{lllllll}
	~~~~\emph{word} & \emph{\#}      & \emph{R}   &  & ~~~~\emph{word} & \emph{\#}    & \emph{R} \\
	~1. return     & 296287 & 0.58 &  & 21. insert   & 8729 & 0.01 \\
	~2. set        & 280213 & 0.93 &  & 22. describe & 8611 & 0.34  \\
	~3. get        & 197072 & 0.45 &  & 23. use      & 8512 & 0.01 \\
	~4. add        & 78769  & 0.88 &  & 24. load     & 8425 & 0.01 \\
	~5. create     & 65044  & 0.69 &  & 25. delete   & 8312 & 0.68 \\
	~6. initialize & 56093  & 0.86 &  & 26. convert  & 8162 & 0.75 \\
	~7. test       & 52116  & 0.68 &  & 27. start    & 7854 & 0.32 \\
	~8. remove     & 35416  & 0.75 &  & 28. clear    & 7828 & 0.55 \\
	~9. check      & 33852  & 0.01 &  & 29. print    & 7507 & 0.77 \\
	10. is         & 20669  & 0.04 &  & 30. find     & 7370 & 0.68 \\
	11. call       & 19459  & 0.03 &  & 31. reset    & 7066 & 0.40 \\
	12. retrieve   & 18035  & 0.33 &  & 32. save     & 7038 & 0.73 \\
	13. update     & 15145  & 0.06 &  & 33. send     & 6950 & 0.77 \\
	14. automate   & 12778  & 0.69 &  & 34. generate & 6936 & 0.48 \\
	15. write      & 12380  & 0.91 &  & 35. close    & 6834 & 0.56 \\
	16. determine  & 12170  & 0.64 &  & 36. compare  & 6684 & 0.55 \\
	17. read       & 9774   & 0.90 &  & 37. indicate & 6434 & 0.55 \\
	18. handle     & 9439   & 0.54 &  & 38. perform  & 6317 & 0.01 \\
	19. to         & 9356   & 0.34 &  & 39. change   & 6081 & 0.01 \\
	20. if         & 8880   & 0.01 &  & 40. show     & 5955 & 0.20
    \end{tabular}
    }
	\caption{The forty most common first words and number of occurrences in the training set of the 2.1m Java method dataset provided by LeClair~\emph{et al.}~\cite{leclair2019neural} (these top forty cover 71\% of the training set).  Column \# is the number of times that word is the first word of a summary in the training set (post stemming).  Column R is the percent of methods with reference summaries with that action word, that a recent baseline~\cite{leclair2019neural} predicted to have that word (this is effectively the recall for that word).} 
	\label{fig:words}
\end{figure}

Figure~\ref{fig:words} shows the top-40 most-common action words that we extracted from the dataset of 2.1m Java methods provided by LeClair~\emph{et al.}~\cite{leclair2019recommendations}.  Our procedure for extracting these words was to use an English language parser to locate the first verb in the sentence, then to stem those verbs to remove conjugation.  (We used the Stanford NLP parser and the NLTK PorterStemmer, our scripts are available in our online appendix, see Section~\ref{sec:appendix}.)  The only exceptions to this procedure were for the words ``is'', ``to'', and ``if.''  The reason for this exception is that there is a very common pattern in code summaries in which those three words broadly classify what methods do, even if they are not typically used as action verbs in English.  Namely, ``is'' tends to be used for code that checks type usage in object oriented programs, ``to'' often refers to code that changes objects from one type to another, and ``if'' often indicates a method checks a particular condition.

It is important to recognize that action words are broad classifications of code behavior rather than rigidly-defined categories.  A scholarly view is that action word prediction is an instance of the \emph{concept assignment problem}~\cite{biggerstaff1993concept} in which words must be used to describe the high level concepts encoded in the low level programming details.  In other words, a programmer can have an intuitive idea of the difference between functions that ``add'' versus ``delete'', and this idea informs what the content of the natural language summary should be.  This intuition cannot be clearly articulated into rules defining the difference.  And yet, code summarization techniques must decide which word to use.

A key point we are trying to make in this paper is that existing code summarization techniques \emph{already} try to predict the action word, and yet action word prediction \emph{on its own} is extremely difficult.  Since almost all code summaries put the action word first, existing tools end up trying to predict this word first.  If the techniques make a poor prediction, the entire summary is likely to be difficult to understand correctly.  Therefore, we advocate for a special emphasis on action word prediction as a stepping stone to code summarization.

Action words are critical to summaries.  A common perception about source code summaries is that they should and most-often do begin with a verb phrase followed by a direct object.  This pattern has been recommended for over 20 years by Kramer~\cite{kramer1999api}, is highlighted by several style guides such as by Google~\cite{GoogleStyleGuide} and Microsoft~\cite{MSStyleGuide}, and has been observed in consistent use in good quality code change documentation~\cite{jiang2017automatically}.

We make a similar observation in both the Java dataset from LeClair et al. and the C/C++ dataset we created. In Java, we found that 94.80\% of method summaries have an action word, of which 56.53\% have one and only one action word. In 80.60\% of comments, an action word is the verb in the first position. In 5.10\% of cases, an action word is in the second position, and in 9.68\% of cases, an action word is in the third position following a simple subject such as ``this method.'' Only 5.60\% did not have an action word in the first, second or third position. And overall in Java, of the summaries with action words, that word is the only verb 54.75\% of the time. So not only are action words perceived as important by style guides and academic literature, programmers actually follow these patterns when writing documentation.

Another observation we make is that some action words are harder to predict than others.  As mentioned above, the decision of which action word to use is a challenge even for humans, who rely on their own intuition rather than strict rules.  Consider column $R$ in Figure~\ref{fig:words}.  That column is the recall of {\small \texttt{ast-attendgru}}, a baseline published at ICSE'19~\cite{leclair2019neural}, when generating summaries over the Java dataset (we extracted these predictions directly from their online reproducibility package and compared them to the reference summaries, after applying stemming to both predictions and references).  While many factors affect prediction and it is important not to take the numbers too literally, it is clear that recall varies considerably for different action words.  For example, recall for ``convert'' is 75\% versus 32\% for ``start'', even though the number of samples is comparable.  The reason, according to experimental evidence by LeClair~\emph{et al.}~\cite{leclair2019neural}, is probably that ``convert'' methods tend to have a more similar structure than ``start'' methods, which helps the model to recognize those methods.  The point is that many action words are very difficult to predict, and if a model cannot chose the action word properly, the entire summary is likely to be incorrect.

In addition, existing models perform far better when they correctly predict the action word.  Consider the table below of BLEU scores for four recent baselines that we trained and tested with both datasets under identical model settings (e.g. vocab size, RNN length, see our online appendix in Section~\ref{sec:appendix} for the implementation).  The column {\small \texttt{default}} is the BLEU score over the whole test set, {\small \texttt{AW c.}} is the BLEU score for just the portion of results when the action word was correctly predicted, {\small \texttt{AW i.}} is when the action word was not correctly predicted.  The column {\small \texttt{AW i.c.}} is a unique setting we created in which we fed the model the correct action word, then let the model predict the rest of the summary (calculated only on the set when the model had incorrectly predicted the action word).  Note that all of these BLEU scores only include the part of the summary \emph{without} the action word, to avoid boosting scores just because the action word is corrected.

\vspace{-0.1cm}
\begin{table}[!h]
	
	\begin{tabular}{p{2.5cm}|lll|l}
		\emph{(Java dataset)} & default & AW c. & AW i. & AW i.c. \\
		attendgru~\cite{iyer2016summarizing}     & 14.63   & 22.21  & 6.16   & 9.35   \\
		ast-attendgru~\cite{leclair2019neural} & 15.29   & 23.15 & 6.36  & 9.32   \\
		graph2seq~\cite{xu2018graph2seq}     & 14.54   & 22.23 & 6.05  & 8.93   \\
		code2seq~\cite{alon2018code2seq}      & 14.73   & 22.55 & 6.16  & 9.08  
	\end{tabular}
\end{table}

\vspace{-0.4cm}

\begin{table}[!h]
	\begin{tabular}{p{2.5cm}|lll|l}
		\emph{(C/C++ dataset)} & default & AW c. & AW i. & AW i.c. \\
		attendgru~\cite{iyer2016summarizing}     & 30.60        & 52.35      & 3.95      & 8.13        \\
		ast-attendgru~\cite{leclair2019neural} & 30.08        & 51.69      & 3.68      & 7.65        \\
		graph2seq~\cite{xu2018graph2seq}     & 22.72        & 43.97      & 2.97      & 6.83        \\
		code2seq~\cite{alon2018code2seq}      & 7.33        & 17.48      & 1.2      & 3.03       
	\end{tabular}
\end{table}
\vspace{-0.1cm}

Performance is roughly 60\% higher when the model predicts the action word correctly.  Good performance is associated with good action word prediction.  Also, performance of the {\small \texttt{AW i.}} set improves 60-100\% if the model is given the correct action word.  The models tend to do much better when the action word is correct regardless of the function -- it is not only that some functions are harder to summarize overall.  The action word is a key factor in generating good summaries.
\section{Experiments}
\label{sec:experiments}

We conduct several experiments exploring action word prediction in the context of existing source code summarization techniques.  We study the effects of different factors on the performance of several baselines.

\subsection{Research Questions}
\label{sec:rqs}

Our major motivation for this paper is to establish action word prediction as a stepping stone towards better source code summarization.  In earlier sections, we discussed the motivation in terms of related literature and evidence that action word prediction is likely to lead to better code summarization.  In this section, we pose the following three Research Questions (RQs) towards studying how well current code summarization approaches perform during action word prediction, and key factors affecting the performance.

\begin{description}
	\item[\textbf{RQ$_1$}] What is the performance of the action word prediction of recent code summarization techniques?
	\vspace{0.1cm}
	\item[\textbf{RQ$_2$}] What is the difference in performance between Java and C/C++ datasets?
	\vspace{0.1cm}
	\item[\textbf{RQ$_3$}] What is the difference in performance under standard and challenge test conditions?
\end{description}

The rationale for RQ$_1$ is that (as we established in the previous section) existing code summarization tools must do action word prediction as part of code summarization anyway, and different approaches are likely to have different performance characteristics on action word prediction.  Since we propose action word prediction as a stepstone problem towards code summarization, we study how well current approaches perform to establish a baseline for further work.  Likewise, the rationale for RQ$_2$ is that different languages tend to have different ``cultures'' and standards for documentation, and it is plausible that these effects are quite large.  Most research in code summarization focuses on one language (usually Java), and the literature is not yet clear on how well results generalize.

The rationale for RQ$_3$ is to quantify how well existing approaches can predict the action word of a summary in the case when only source code structure is available.  Source code summarization, like many program comprehension research problems, is highly dependent on the words used in the source code.  LeClair~\emph{et al.} found that BLEU scores approximately halve when words are not available and the tools must rely solely on code structure~\cite{leclair2019neural}.  They term the situation when only code structure is available as a ``challenge'' experiment, in contrast to a ``standard'' experiment when both structure and words from code are available.  However, as pointed out in Section~\ref{sec:bg}, an increasing number of research efforts are finding that code structure significantly improves code summarization.  Evidence from Alon~\emph{et al.}~\cite{alon2019code2seq}, LeClair~\emph{et al.}~\cite{leclair2019neural, leclair2018adapting}, and Haque~\emph{et al.}~\cite{haque2020improved} all point to how code structure seems to help code summarization tools broadly classify code behavior, even if good internal documentation is required to produce detailed summaries.  We surmise that a way structure manifests itself in summaries is via the action word, and therefore ask RQ$_3$.

\subsection{Methodology}
\vspace{-0.1cm}

Our methodology for answering our research questions is, essentially, to train each baseline over each dataset, then use established quantitative metrics to measure the difference in performance over the test set for that dataset.  This is the typical evaluation procedure used in a vast majority of source code summarization papers, except modified slightly to measure the action word prediction instead of the entire summary.

Note that we use a data-driven experiment design rather than a human study.  While it is tempting to suggest that a human study is always superior, in fact for comparison in this paper a data-driven evaluation has several advantages.  The situation is akin to machine translation around the late 1990s: the introduction of data-driven evaluation procedures and metrics (such as BLEU~\cite{Papineni:2002:BMA:1073083.1073135}) led to a revolution of progress because it was possible to consistently evaluate tens of thousands of examples.  Human studies certainly have the advantage of allowing for deep subjective evaluation, but there are also two major disadvantages: 1) including humans restricts the amount of output that can be evaluated to perhaps a few hundred examples, and 2) humans are subject to the vicissitudes of nature e.g. fatigue and biases, which means that the results are not reproducible.  That is why almost all code summarization papers also use data-driven evaluations -- the ability to quickly evaluate large datasets and reproduce those findings under controlled settings is an important benefit to overall progress.

Our training procedure is to train each baseline for a maximum of 10 epochs or 30 hours, and select the model that achieved the highest validation accuracy.  The code2seq baseline running on the C/C++ dataset was the only approach to timeout at 30 hours.  All other configurations (including code2seq for Java) completed 10 epochs within 30 hours.

\subsubsection{Baselines}
\label{sec:baselines}

While there is a plethora of available code summarization techniques (see Section~\ref{sec:bg}), we choose five that are broadly representative of different families of approaches.  Namely, seq2seq-like approaches that treat code as text, approaches that flatten the AST in the encoder, tree and graph neural network solutions, and path-based approaches.  In all cases, we reimplement the approaches using a framework provided by Haque~\emph{et al.}~\cite{haque2020improved}.  We used their framework in order to minimize experimental variables -- we do not use individual implementations provided in different reproducibility packages.  The reason is that many factors change such as RNN length, vocab size, dataset preprocessing, etc.  To ensure veracity of the results, we use a single experimental framework for implementation to control these factors.  See our online appendix for full implementations (Section~\ref{sec:appendix}).

{\small \textbf{\texttt{attendgru}}}  This baseline is representative of seq2seq-like approaches as proposed by Iyer~\emph{et al.}~\cite{iyer2016summarizing}.  This is the only approach that does not explicitly include the AST, so this baseline is not applicable to the challenge experiments.

{\small \textbf{\texttt{ast-attendgru}}}  This baseline represents approaches that flatten the AST into a sequence, then use a seq2seq-like approach to create the summary.  Our implementation is from LeClair~\emph{et al.}~\cite{leclair2019neural} but is closely-related to other flat AST techniques e.g. Hu~\emph{et al.}~\cite{hu2018deep}.

{\small \textbf{\texttt{ast-attendgru-fc}}}  Haque~\emph{et al.}~\cite{haque2020improved} proposed an extension to code summarization tools that includes ``file context'', which they define as all the other functions in the same file as the function being summarized.  This baseline is the best-performing solution from their experiments.  It is identical to {\small {\texttt{ast-attendgru}}} except with their ``file context encoder.''


{\small \textbf{\texttt{graph2seq}}}  Allamanis~\emph{et al.}~\cite{allamanis2018learning} use a graph neural network (GNN) to model the AST.  Their paper focuses on code generation, but suggest that it is possible to use the GNN encoder for code summarization as well.  Haque~\emph{et al.}~\cite{haque2020improved} provide an implementation based on graph2seq by Xu~\emph{et al.}~\cite{xu2018graph2seq}, and we use this implementation as a baseline.

{\small \textbf{\texttt{code2seq}}}  Alon~\emph{et al.}~\cite{alon2019code2seq, alon2019code2vec} are pursuing path-based encoding solutions that, in short, randomly select 100-200 paths in the AST and use an attention mechanism to attend to different paths for different words.  We use their approach as a representative of path-based solution. 

This list of baselines is not an exhaustive list of approaches from all papers.  We instead pick one approach representative of different families of approaches.  However, readers may note that we did not include approaches that focus on novel training procedures or optimizers (e.g. approaches that use combined reinforcement and deep learning~\cite{wan2018improving}).  There are two reasons for this decision.  First and foremost, those approaches use RL or other techniques to help train the model on entire summaries, and since summaries are almost always predicted one word at a time, the benefit almost entirely appears when writing the latter parts of the summaries.  Second, those approaches introduce a large number of additional experimental variables which are difficult to control.  We did not use a Transformer-based approach because Haque~\emph{et al.}~\cite{haque2020improved} report that low performance from Transformer models (however, just as this paper was going to review, Ahmad~\emph{et al.}~\cite{ahmad2020transformer} released a Transformer-based model accepted to ACL'20.).

\subsubsection{Metrics}  We use precision, recall, and F-measure to evaluate the performance for action word prediction~\cite{powers2011evaluation}.  We also use confusion matrices to visualize performance across difference words.  Essentially what we do is treat each action word as a class into which a subroutine can be placed.  So for example, a function may be a ``get'' subroutine, an ``insert'' subroutine, etc.  Note that this is different than code summarization, where typically BLEU or ROUGE scores are used to evaluate a whole sentence -- we only use BLEU when we have a full sentence to evaluate.

\subsubsection{Datasets}

We create two action word prediction datasets from two code summarization datasets.  First, as mentioned above, we use a Java dataset of 2.1m method/summary pairs for code summarization provided by LeClair~\emph{et al.}~\cite{leclair2019recommendations}.  Second, we create our own C/C++ dataset of 1.1m function/summary pairs, by downloading 18,775 C/C++ projects from GitHub and following LeClair~\emph{et al.}'s recommendations from NAACL'19 for code summarization datasets (e.g. splitting by project, removing auto-generated code, minimum/maximum summary lengths, and other quality filters).  We use a model published by Eberhart~\emph{et al.}~\cite{eberhart2019automatically} to extract the summary from comments for C/C++ functions, since C/C++ tends not to have the same rules for writing subroutine summary documentation as Java.

Two steps were necessary to convert the code summarization datasets to action word prediction datasets.  First, we extract the action word from each summary by using the Stanford NLP package~\cite{manning2014stanford}, except for summaries that start with the word ``is'', ``to'', or ``if'', for reasons discussed in Section~\ref{sec:awp}.  Second, we used the NLTK PorterStemmer on each action word because of the large number of patterns such as ``initializer for ...'' and ``initializes ...''  Stemming has the effect of reducing the vocab size by grouping words together, which is different from code summarization in which stemming is rare.  

\subsection{Threats to Validity}

Like all experiments, this paper carries threats to the validity of the results.  Three threats loom the largest:  One threat is that the data may not be representative.  We attempt to mitigate this risk by using large datasets in different languages, but the reality is that different data may lead to different results.  Another threat is that the scripts we use to extract action words, stem words, etc., do not have perfect accuracy.  We use community best practices to minimize errors, and release all scripts online vote public vetting, but again, results may differ under different conditions.  The datasets are simply too large to allow for large-scale manual inspection. Finally, we did not test if masking another word, for example a noun instead of the verb, gives the same results. We believe that action words are very specific and cannot be replaced with generic verbs to convey the same meaning, but nouns can be replaced with pronouns.

\section{Experimental Results}

We present our experimental results in the form of answers to each research question from the previous section.  To condense results and make the key points understandable in the paper, we present results using four settings: \textbf{top-40}, \textbf{top-10}, \textbf{top-10n}, and \textbf{get/set}.  The top-40 setting means that the models attempt to predict the forty most-common action words, or ``other'' if predicted ot be a less-common action word.  The forty most-common action words covers over 71\% of the training set and aligns with the discussion in Section~\ref{sec:awp}.  The top-10 setting is similar except that it covers only the ten most-common action words, and is a convenient setting for presentation due to paper formatting requirements. The Top-10n is \emph{next} ten most-common results without get, set, and return (examples in this category go into the other category).  Since the dataset is somewhat unbalanced towards these top-3 action words, it is plausible that strong performance in these categories could outweigh poor performance elsewhere.  So, we present top-10n to show performance on other categories.




The get/set setting is a diagnostic setting focusing on get and set examples only. There is no large ``other" category. The purpose of this setting is to show the limits of the model in an ``easy'' situation: plenty of training data and clear conceptual difference between the two categories. It is not meant as a realistic situation.  The philosophical intent is to create an experimental setting that we, the research community, should be able to solve with extremely high accuracy -- if we cannot even distinguish gets and sets words, then we have little hope of predicting an entire code summary. On the other hand, the get/set diagnostic problem can serve as a stepping stone: models that solve it with high accuracy may serve as a foundation for better code summarization techniques.  The situation is akin to early computer vision models designed to distinguish e.g. cat from trucks.  The idea is a diagnostic problem that should not be confused for the real world.

\vspace{-0.09cm}
\subsection{RQ$_1$: Baseline Performance}

In this section, we answer RQ$_1$ in the context of the baseline performance over the Java dataset in standard conditions. 
The lower area of Table~\ref{tab:rq1} shows the precision/recall values for the baselines in the four settings described above.  The best overall performer is {\small \texttt{ast-attendgru-fc}}.  Yet, the difference over {\small \texttt{attendgru}} and {\small \texttt{ast-attendgru}} is not high despite reports from Haque~\emph{et al.}~\cite{haque2020improved} that significant improvement is possible in terms of BLEU score over the entire code summary.  Our interpretation is that the benefit that {\small \texttt{ast-attendgru-fc}} provides is concentrated in the later portions of the summary -- the action word prediction is similar in all three models.


{\setlength\extrarowheight{5pt}\centering\setlength\tabcolsep{2pt}
	\begin{table}[!b]
		\vspace{-0.1cm}
		\begin{tabular}{rlllllllllll}
			\multicolumn{1}{l}{\vspace{-0.5cm}\multirow{2}{*}{\textbf{Java / stan.}}} &        &     &     &     &        &            &      &        &       &    \\
			\multicolumn{1}{l}{top-10}                                 & \rotatebox{90}{return} & \rotatebox{90}{set} & \rotatebox{90}{get} & \rotatebox{90}{add} & \rotatebox{90}{create} & \rotatebox{90}{initialize} & \rotatebox{90}{test} & \rotatebox{90}{remove} & \rotatebox{90}{check} & \rotatebox{90}{is} &\rotatebox{90}{\emph{other}} \\
			return                                               &7748&25   &1144&11  &70  &11  &1   &5   &78 &1 &1927 \\
			set                                                  &18  &10356&10  &40  &12  &17  &2   &1   &0  &0 &583 \\
			get                                                  &5202&19   &2595&6   &11  &9   &3   &1   &8  &0 &844 \\
			add                                                  &4   &44   &2   &2925&14  &1   &10  &2   &0  &0 &321 \\
			create                                               &80  &21   &18  &40  &1015&14  &17  &0   &2  &0 &496 \\
			initialize                                           &31  &21   &2   &2   &23  &1528&0   &0   &0  &0 &290 \\
			test                                                 &67  &2    &2   &1   &1   &1   &1025&3   &23 &0 &381 \\
			remove                                               &6   &2    &1   &1   &0   &2   &3   &1132&0  &0 &354 \\
			check                                                &393 &8    &5   &2   &3   &0   &52  &3   &343&0 &657 \\
			is                                                   &134 &36   &6   &15  &11  &5   &3   &6   &21 &31 &366 \\
			\emph{other}                                                   &3444 &2123   &617   &493  &625  &159   &549   &172   &211 &4 &22420 \\
		\end{tabular}
	\end{table}
	\vspace{0.1cm}
}

{\setlength\tabcolsep{3pt}
\begin{table}[!b]
	\begin{tabular}{lllllllllllll}
		\multirow{2}{*}{\textbf{Java / stan.}}      & \multicolumn{3}{c}{top-40}    & \multicolumn{3}{c}{top-10}    & \multicolumn{3}{c}{top-10n}   & \multicolumn{3}{c}{get/set} \\
		& ~p & ~r & ~f                     & ~p & ~r & ~f                     & ~p & ~r & ~f                     & ~p       & ~r       & ~f       \\
		\multicolumn{1}{l|}{attendgru}        &.54&.45& \multicolumn{1}{l|}{.46} &.68&.63& \multicolumn{1}{l|}{.61} &.71&.52& \multicolumn{1}{l|}{.53} &.99&.99&.99         \\
		\multicolumn{1}{l|}{ast-attendgru}    &.59&.44& \multicolumn{1}{l|}{.46} &.67&.62& \multicolumn{1}{l|}{.62} &.70&.53& \multicolumn{1}{l|}{.53} &.99&.99&.99         \\
		\multicolumn{1}{l|}{ast-attendgru-fc} &.53&.47& \multicolumn{1}{l|}{.46} &.70&.61& \multicolumn{1}{l|}{.61} &.73&.50& \multicolumn{1}{l|}{.52} &.99&.99&.99         \\
		\multicolumn{1}{l|}{graph2seq}        &.53&.47& \multicolumn{1}{l|}{.47} &.68&.62& \multicolumn{1}{l|}{.62} &.70&.53& \multicolumn{1}{l|}{.54} &.99&.99&.99         \\
		\multicolumn{1}{l|}{code2seq}        &.55&.45& \multicolumn{1}{l|}{.47} &.67&.61& \multicolumn{1}{l|}{.61} &.71&.53& \multicolumn{1}{l|}{.54} &.99&.99&.99           
	\end{tabular}
	\vspace{0.1cm}
	\caption{\emph{(top)} Confusion matrix showing results for top-10 action words for ast-attendgru-fc, the best performer in terms of f-measure.  \emph{(bottom)} Overall results under standard conditions in the Java dataset.}
	\label{tab:rq1}
\end{table}
}

\vspace{-0.125cm}
A possible explanation for this similar performance is that all three models rely on a recurrent network (a GRU) encoder of the function's text information, moderated by an attention mechanism.  It is likely that the attention mechanism learns to look at the first word of the function name when predicting the first word of the summary.  In fact, in the Java dataset, we found that 43\% of the methods have the summary's action word within the first four words in the method's signature (e.g. the verb ``convert'' for the method ``convertMp3toWav'') and 67\% have the action word somewhere in the method.

In contrast, {\small \texttt{graph2seq}} had much lower performance, despite competitive BLEU scores when creating a whole summary (see~\cite{haque2020improved, leclair2020improved}, plus Section~\ref{sec:awp}).  The reason may be explained by recent results by LeClair~\emph{et al.}~\cite{leclair2020improved}: the graph neural network design excels at choosing related next words in a summary by exploiting the graph's connections, but can actually perform worse when these connections are not available -- significant configuration is necessary and competitive performance cannot be expected out of the box.  

The confusion matrix at the top of Table~\ref{tab:rq1} shows that top-10 results for Java under standard conditions for {\small \texttt{ast-attendgru-fc}}.  In general, a clear centerline is visible as the model correctly predicts a majority of the action words.  One observation is that the model tends to confuse action words that are conceptually similar more often than conceptually distinct ones.  For example, ``get'' and ``return'' are often interchangeable, so it is unsurprising that a significant number of the errors are from mistaking just these two words.  Also, words like check/test and create/initialize tend to be confused.  In contrast, the model make very few errors for words with distinct meanings such as add/remove and get/set.  The model had little trouble distinguishing get and set under standard conditions in Java.

{\setlength\extrarowheight{5pt}\centering\setlength\tabcolsep{2pt}
	\begin{table}[!b]
		\vspace{-0.1cm}
		\begin{tabular}{rlllllllllll}
			\multicolumn{1}{l}{\vspace{-0.5cm}\multirow{2}{*}{\textbf{C/C++ / stan.}}} &        &     &     &     &        &            &      &        &       &    \\
			\multicolumn{1}{l}{top-10}                                 & \rotatebox{90}{return} & \rotatebox{90}{get} & \rotatebox{90}{set} & \rotatebox{90}{check} & \rotatebox{90}{call} & \rotatebox{90}{initialize} & \rotatebox{90}{is} & \rotatebox{90}{read} & \rotatebox{90}{create} & \rotatebox{90}{add} &\rotatebox{90}{\emph{other}} \\
			return                                               &926  &79  &2  &32  &6  &4  &7  &18  &5  &4 &912   \\
			get                                                  &50  &529  &3  &4  &4  &1  &10  &9  &2  &0 &381   \\
			set                                                  &7  &5  &597  &0  &7  &8  &11  &0  &1  &1 &506   \\
			check                                                  &9  &2  &1  &333  &0  &2  &8  &0  &0  &1 &253   \\
			call                                               &0  &2  &7  &0  &472  &4  &4  &0  &2  &2 &228  \\
			initialize                                           &6  &1  &4  &0  &6  &489  &2  &0  &6  &0 &233   \\
			is                                                 &21  &8  &7  &9  &4  &5  &469  &1  &1  &1 &458   \\
			read                                               &7  &9  &1  &0  &1  &0  &0  &362  &1  &0 &190   \\
			create                                                &6  &1  &0  &0  &2  &13  &2  &0  &286  &1 &209   \\
			add                                                   &3  &2  &0  &0  &3  &3  &2  &0  &2  &211 &139 \\
			\emph{other}                                                   &479  &334  &175  &177  &216  &167  &146  &128  &95  &112 &34759
		\end{tabular}
	\end{table}
	\vspace{0.1cm}
}

{\setlength\tabcolsep{3pt}
\begin{table}[!b]
	\begin{tabular}{lllllllllllll}
		\multirow{2}{*}{\textbf{C/C++ / stan.}}      & \multicolumn{3}{c}{top-40}    & \multicolumn{3}{c}{top-10}    & \multicolumn{3}{c}{top-10n}   & \multicolumn{3}{c}{get/set} \\
		& ~p & ~r & ~f                     & ~p & ~r & ~f                     & ~p & ~r & ~f                     & ~p       & ~r       & ~f       \\
		\multicolumn{1}{l|}{attendgru}        &.70&.55& \multicolumn{1}{l|}{.60} &.71&.63& \multicolumn{1}{l|}{.67} &.73&.63& \multicolumn{1}{l|}{.68} &.96&.96&.96 \\
		\multicolumn{1}{l|}{ast-attendgru}    &.69&.55& \multicolumn{1}{l|}{.60} &.72&.61& \multicolumn{1}{l|}{.66} &.75&.62& \multicolumn{1}{l|}{.67} &.96&.97&.96 \\
		\multicolumn{1}{l|}{ast-attendgru-fc} &.65&.55& \multicolumn{1}{l|}{.58} &.68&.60& \multicolumn{1}{l|}{.63} &.74&.55& \multicolumn{1}{l|}{.63} &.96&.96&.96 \\
		\multicolumn{1}{l|}{graph2seq}        &.70&.49& \multicolumn{1}{l|}{.55} &.71&.60& \multicolumn{1}{l|}{.65} &.77&.57& \multicolumn{1}{l|}{.65} &.95&.95&.95 \\
		\multicolumn{1}{l|}{code2seq}         &.64&.41& \multicolumn{1}{l|}{.47} &.66&.50& \multicolumn{1}{l|}{.56} &.67&.56& \multicolumn{1}{l|}{.60} &.96&.96&.96     
	\end{tabular}
	\vspace{0.1cm}
	\caption{\emph{(top)} Confusion matrix showing results for top-10 action words for ast-attendgru-fc, the best performer in terms of f-measure.  \emph{(bottom)} Overall results under standard conditions in the C/C++ dataset.}
\end{table}
}

\newpage

\vspace{0.1cm}
\subsection{RQ$_2$: Comparing Java and C/C++}
\vspace{0.09cm}

In this section, we compare Java and C/C++ results.  Note that our objective is to start to understand the behavior of the baselines in these datasets related to the data itself.  We do not claim that one dataset is ``better'' than another from these results.  Instead, we observe that the baselines almost all achieved higher levels of performance in C/C++, which seems to be related to two factors.  First, the top-$n$ action words are different in each language, and even words that are the same have slightly different meanings.  The word ``initialize'', for instance, tends to refer to a specific kind of memory allocation and initialization in C/C++, compared to Java in which it tends be used more generally, such as to refer to setting up data in an arbitrary class structure.  The composition of the words is also slightly different.  For example, the word ``call'' appears in C/C++, and the models make very few errors when detecting this word because a vast majority of caller functions have the word call in their name.

Second, errors in the Java dataset are somewhat concentrated in the word ``return.''  One possible explanation is a cultural tendency among Java programmers to write summaries in the form ``returns ...'' instead of an action word in some cases.  For example, Java programmers seem to be more likely to write ``returns whether ...'' instead of ``checks ...'', which seems to lead to a significant number of errors in confusing ``check'' for ``return.''  Since the training and test sets are derived from examples of human-written summaries, any model is likely to learn these errors.  Note that we do not recommend a specific remedy, other than to be aware that these factors do exist in datasets of source code summarization and may require adjustments based on the model and use case.

{\setlength\tabcolsep{8pt}
	\begin{table}[!b]\centering
		\vspace{-0.2cm}
		\begin{tabular}{lllllll}
			\multirow{2}{*}{\textbf{Java / stan.}} & \multicolumn{3}{c}{original dataset}   & \multicolumn{3}{c}{1m dataset}     \\
			& ~p & ~r & ~f        & ~p & ~r & ~f        \vspace{0.1cm}               \\
			\multicolumn{1}{l|}{attendgru}        &.68&.63& \multicolumn{1}{l|}{.61} &.68&.61& .60       \\
			\multicolumn{1}{l|}{ast-attendgru}    &.67&.62& \multicolumn{1}{l|}{.62} &.69&.61& .61      \\
			\multicolumn{1}{l|}{ast-attendgru-fc} &.70&.61& \multicolumn{1}{l|}{.61} &.70&.58& .59       \\
			\multicolumn{1}{l|}{graph2seq}        &.68&.62& \multicolumn{1}{l|}{.62} &.67&.62& .61       \\
			\multicolumn{1}{l|}{code2seq}         &.67&.61& \multicolumn{1}{l|}{.61} &.67&.62& .61           
		\end{tabular}
		\label{tab:Java1mstandard}
	\end{table}
	\vspace{0.0cm}
	\begin{table}[!b]\centering
		\vspace{0.0cm}
		\begin{tabular}{lllllll}
			\multirow{2}{*}{\textbf{Java / chal.}}       & \multicolumn{3}{c}{original dataset}   & \multicolumn{3}{c}{1m dataset}   \\
			& ~p & ~r & ~f       & ~p & ~r & ~f       \vspace{0.1cm}             \\
			\multicolumn{1}{l|}{attendgru}        &.04&.09& \multicolumn{1}{l|}{.05} &.04&.09& .05         \\
			\multicolumn{1}{l|}{ast-attendgru}    &.60&.33& \multicolumn{1}{l|}{.36} &.55&.33& .34        \\
			\multicolumn{1}{l|}{ast-attendgru-fc} &.54&.43& \multicolumn{1}{l|}{.45} &.50&.51& .43     \\
			\multicolumn{1}{l|}{graph2seq}        &.34&.14& \multicolumn{1}{l|}{.13} &.25&.15& .14       \\
			\multicolumn{1}{l|}{code2seq}         &.24&.12& \multicolumn{1}{l|}{.11} &.17&.12& .10         
		\end{tabular}
		\vspace{0.1cm}
		\caption{Overall results of predicting the top-10 action words under standard \emph{(top)} and challenge \emph{(bottom)} conditions in the original Java dataset vs 1m dataset.}
		\label{tab:Java1mchallenge}
	\end{table}
}

We also note that the differences between Java and C/C++ cannot be explained merely by the different size of the dataset.  In any neural network-based solution, a question will arise as to how many examples are necessary to train the model.  Answering this question depends on the diversity of the dataset, training configuration settings such as learning rate, and details of the model itself such as the architecture (e.g. RNN, CNN, FCN) and number of model parameters.  Nevertheless, in very similar environments, different performance would be observed with different numbers of training examples.  In our case, we have set the model and training configuration to be identical for Java and C/C++ experiments.  We may surmise that the Java and C/C++ datasets are different enough to produce different results due to the content of the datasets (e.g. word usage, code structure differences, or other factors), but it is still possible that performance differences can be explained by the much larger number of examples in the Java dataset (2m examples) versus the C/C++ dataset (1m examples).

To mitigate dataset size as a factor, we created a subset of the Java dataset of equal size to the C/C++ dataset.  We created this dataset by randomly selecting examples from the full Java training set.  The test set is identical.  We denote this dataset as the ``1m'' dataset in Table~\ref{tab:Java1mchallenge}.  The ``original dataset'' in the table are results of the top-10 action word prediction using the default 2m Java dataset (note these are duplicated from the top-10 column in Table~\ref{tab:rq1} for convenience).  The ``1m dataset'' are the top-10 results for the baselines when training only on the 1m dataset and testing on the same test set as the original Java dataset.  What we observe is that very few of the results change on the standard and challenge conditions.  The 1m dataset is about 1\% lower in terms of f-score for the standard set, and 1-2\% lower for the challenge set.

We conclude that we did not find evidence that dataset size alone is the key factor in the difference between Java and C/C++.  The most likely explanation lies in the differences in code itself, such as a culture of longer identifier names and more internal documentation in Java~\cite{kramer1999api}.

\subsection{RQ$_3$: Challenge Conditions}

Recall that challenge conditions are those in which all textual information in the source code is removed; only the structure of the code is available to the model for training and testing~\cite{leclair2019neural}.  Challenge conditions are something of a holy grail for code summarization because successful summarization under challenge conditions would mean that even totally undocumented code could be described using automated techniques.  Action word prediction under challenge conditions is a strong starting point because the action word is very important to the summary overall.

The highest level of performance we observed was for the get/set condition in both Java and C/C++.  In the Java dataset, the {\small \texttt{ast-attendgru}} achieved 96\% precision/recall.  It achieved 77\% precision and recall for C/C++.  We attribute these high values to predictable structure of getters and setters in both languages: getters tend to have return data and no parameters, while setters have parameters and no return data.  Another observation is that context from other functions seems to have a strong positive impact on results under challenge conditions -- {\small \texttt{ast-attendgru-fc}} routinely obtains the highest performance and is the only baseline to consider contextual information.  The implication for future work is that if a function is not described well, it may still be possible to predict the action word for that function's summary by looking at the other functions in the same file.  The model may be able to learn, for example, that if a ``read'' is present, a ``write'' is likely, or that if several other functions in the same file include the word ``initialize'', then this function is also likely to use it.  Performance levels in the challenge set are otherwise low. 


{\setlength\extrarowheight{5pt}\centering\setlength\tabcolsep{2pt}
	\begin{table}[!b]
		\vspace{-0.2cm}
			\begin{tabular}{rlllllllllll}
				\multicolumn{1}{l}{\vspace{-0.5cm}\multirow{2}{*}{\textbf{Java / chal.}}} &        &     &     &     &        &            &      &        &       &    \\
				\multicolumn{1}{l}{top-10}                                 & \rotatebox{90}{return} & \rotatebox{90}{set} & \rotatebox{90}{get} & \rotatebox{90}{add} & \rotatebox{90}{create} & \rotatebox{90}{initialize} & \rotatebox{90}{test} & \rotatebox{90}{remove} & \rotatebox{90}{check} & \rotatebox{90}{is} &\rotatebox{90}{\emph{other}} \\
				return                                               &2961&23   &3340&2  &17 &59  &3   &0 &11&0 &4605   \\
				set                                                  &11  &8107 &14  &33 &3 &6  &3   &0 &0 &0 &2862   \\
				get                                                  &1914&23   &3386&1  &35 &53  &2   &0  &2&0 &3282    \\
				add                                                  &14  &179  &0  &801&2 &4  &2   &6&0 &0 &2315   \\
				create                                               &68  &15   &6  &1  &48&27  &11  &0  &0 &1 &1526   \\
				initialize                                           &12  &40   &4   &0   &3  &1427&5   &0  &0 &0 &406   \\
				test                                                 &38  &1    &10  &1   &0  &5   &567&0  &3 &0 &881   \\
				remove                                               &24  &36   &2  &23 &0  &0   &1   &19&0 &0 &1396   \\
				check                                                &206 &6   &79  &1   &1  &1   &21  &0 &19&0 &1132   \\
				is                                                   &53  &30   &61  &0   &0  &1   &2   &0  &1 &5 &481  \\
				\emph{other}                                                   &1561  &1705   &1096  &107   &79~~  &131   &212   &12~~  &14~~ &0~~ &25900   
		\end{tabular}
	\end{table}
	\vspace{0.0cm}
	\begin{table}[!b]
	\vspace{-0.1cm}
	\begin{tabular}{rlllllllllll}
		\multicolumn{1}{l}{\vspace{-0.5cm}\multirow{2}{*}{\textbf{C/C++ / chal.}}} &        &     &     &     &        &            &      &        &       &    \\
		\multicolumn{1}{l}{top-10}                                 & \rotatebox{90}{return} & \rotatebox{90}{get} & \rotatebox{90}{set} & \rotatebox{90}{check} & \rotatebox{90}{call} & \rotatebox{90}{initialize} & \rotatebox{90}{is} & \rotatebox{90}{read} & \rotatebox{90}{create} & \rotatebox{90}{add} &\rotatebox{90}{\emph{other}} \\
		return                                               &124~~  &9~~~  &0~~~  &4~~~  &0~~~  &0~~~  &0~~~  &0~~~  &0~~~  &0~~~ &1858   \\
		get                                                  &25  &10  &0  &1  &0  &0  &0  &0  &0  &0 &957   \\
		set                                                  &1  &0  &3  &0  &0  &0  &0  &0  &0  &0 &1139   \\
		check                                                  &3  &1  &0  &10  &0  &0  &8  &0  &0  &0 &592   \\
		call                                               &0  &0  &0  &0  &1  &1  &0  &0  &0  &0 &719  \\
		initialize                                           &0  &0  &0  &0  &0  &12  &0  &0  &1  &0 &734   \\
		is                                                 &1  &1  &0  &1  &0  &0  &4  &0  &0  &0 &977   \\
		read                                               &1  &0  &0  &2  &0  &0  &0  &1  &0  &0 &567   \\
		create                                                &0  &0  &0  &0  &0  &1  &0  &0  &4  &0 &515   \\
		add                                                   &0  &0  &0  &0  &0  &1  &0  &0  &0  &0 &364 \\
		\emph{other}                                                   &76  &10  &2  &11  &0  &11  &0  &3  &12  &0 &36663
	\end{tabular}
\end{table}
}

{\setlength\tabcolsep{3pt}
\begin{table}[!b]
	\begin{tabular}{lllllllllllll}
		\multirow{2}{*}{\textbf{Java / chal.}}      & \multicolumn{3}{c}{top-40}    & \multicolumn{3}{c}{top-10}    & \multicolumn{3}{c}{top-10n}   & \multicolumn{3}{c}{get/set} \\
		& ~p & ~r & ~f                     & ~p & ~r & ~f                     & ~p & ~r & ~f                     & ~p       & ~r       & ~f       \\
		\multicolumn{1}{l|}{attendgru}        &.01&.02& \multicolumn{1}{l|}{.01} &.04&.09& \multicolumn{1}{l|}{.05} &.07&.09& \multicolumn{1}{l|}{.08} &.29&.50&.36 \\
		\multicolumn{1}{l|}{ast-attendgru}    &.26&.12& \multicolumn{1}{l|}{.12} &.60&.33& \multicolumn{1}{l|}{.36} &.36&.22& \multicolumn{1}{l|}{.24} &.96&.96&.96 \\
		\multicolumn{1}{l|}{ast-attendgru-fc} &.35&.20& \multicolumn{1}{l|}{.23} &.54&.43& \multicolumn{1}{l|}{.45} &.43&.27& \multicolumn{1}{l|}{.30} &.97&.97&.97 \\
		\multicolumn{1}{l|}{graph2seq}        &.09&.04& \multicolumn{1}{l|}{.04} &.34&.14& \multicolumn{1}{l|}{.13} &.26&.09& \multicolumn{1}{l|}{.08} &.82&.74&.74 \\
		\multicolumn{1}{l|}{code2seq}         &.04&.03& \multicolumn{1}{l|}{.03} &.24&.12& \multicolumn{1}{l|}{.11} &.18&.09& \multicolumn{1}{l|}{.08} &.79&.78&.78 
	\end{tabular}

	\vspace{0.3cm}
	\begin{tabular}{lllllllllllll}
	\multirow{2}{*}{\textbf{C/C++ / chal.}}      & \multicolumn{3}{c}{top-40}    & \multicolumn{3}{c}{top-10}    & \multicolumn{3}{c}{top-10n}   & \multicolumn{3}{c}{get/set} \\
		& ~p & ~r & ~f                     & ~p & ~r & ~f                     & ~p & ~r & ~f                     & ~p       & ~r       & ~f       \\
	\multicolumn{1}{l|}{attendgru}        &.02&.02& \multicolumn{1}{l|}{.02} &.07&.09& \multicolumn{1}{l|}{.08} &.08&.09& \multicolumn{1}{l|}{.08} &.27&.50&.35 \\
	\multicolumn{1}{l|}{ast-attendgru}    &.51&.07& \multicolumn{1}{l|}{.10} &.51&.10& \multicolumn{1}{l|}{.10} &.08&.09& \multicolumn{1}{l|}{.08} &.77&.77&.77 \\
	\multicolumn{1}{l|}{ast-attendgru-fc} &.47&.14& \multicolumn{1}{l|}{.20} &.49&.20& \multicolumn{1}{l|}{.26} &.66&.12& \multicolumn{1}{l|}{.15} &.82&.81&.81 \\
	\multicolumn{1}{l|}{graph2seq}        &.02&.02& \multicolumn{1}{l|}{.02} &.16&.09& \multicolumn{1}{l|}{.08} &.08&.09& \multicolumn{1}{l|}{.09} &.54&.53&.51 \\
	\multicolumn{1}{l|}{code2seq}         &.02&.02& \multicolumn{1}{l|}{.02} &.07&.09& \multicolumn{1}{l|}{.08} &.08&.09& \multicolumn{1}{l|}{.09} &.56&.56&.56        
	\end{tabular}

	\vspace{0.0cm}
	\caption{\emph{(top)} Confusion matrices showing results for top-10 action words for ast-attendgru, which in challenge conditions uses only the AST.  \emph{(bottom)} Overall results under challenge conditions in the Java and C/C++ datasets.}
	\label{tab:chalmatrices}
	\vspace{-0.15cm}
\end{table}
}

Another observation we make in the challenge set is that the top-10 performance is higher for some models than the top-10n performance.  This behavior is opposite of the standard set.  In the standard set, many errors were concentrated in confusion between get and return.  In contrast, under challenge conditions the baselines tend to place many functions into the ``other'' category, except for a few return/get/set functions that they predict correctly (see the top-10 C/C++ challenge confusion matrix in Table~\ref{tab:chalmatrices}).  The {\small \texttt{ast-attendgru-fc}} model is the exception, which we attribute to that model's access to words in the file context.  The same model without access to file context ({\small \texttt{ast-attendgru}}) achieved much lower performance.

The differences between the Java and C/C++ datasets are evident in the challenge results.  Consider the top-10 results for {\small \texttt{ast-attendgru}} and {\small \texttt{ast-attendgru-fc}}.  In the Java dataset, the {\small \texttt{ast-attendgru-fc}} is about 25\% higher in terms of F-measure (36\% to 45\%).  But in the C/C++ dataset, the performance is well over double (10\% to 26\%) in terms of F-measure.  Almost all of this improvement is attributable to increased recall in the get/set/return functions -- that is, the model does a better job of finding these instead of placing them in the other category.  This difference is borne out in the top-10n results as well.  The difference in Java is about 25\%, but is nearly double in C/C++.  For the top-10n, most of the improvement in F-measure is due to greatly increased precision (8\% to 66\%).  The improved precision is because {\small \texttt{ast-attendgru-fc}} avoids misplacing several functions into the other category, though recall remains quite low because a majority of functions still end up in other.  The takeaway is that context helps when it is available, even in the extreme conditions of the challenge set, but existing models are still far away from real-world usability.

\begin{figure}[b!]
	\vspace{-0.4cm} 
	{\small	
		\begin{tabular}{lm{5cm}}
			\emph{reference}   & lookup chunk type debug name \\
			\emph{model output} & return a pointer to the chunk of chunk \\
			\emph{$\hookrightarrow$ given correct a.w.}   & lookup chunk type debug name \\
		\end{tabular}
		\begin{tabular}{lm{5cm}}
			\emph{context}      & \begin{scriptsize}
				\begin{verbatim}
				const char sctp cname const sctp subtype
				t cid if cid chunk 0 return illegal chunk <TRUNC>\end{verbatim}\end{scriptsize} \\
			\vspace{0.9cm}\emph{raw}      & \vspace{-0.4cm}\begin{scriptsize}\begin{verbatim}const char *sctp_cname(const sctp_subtype_t cid)
				{
				if (cid.chunk < 0)
				return "illegal chunk	<TRUNC>\end{verbatim}\end{scriptsize}
		\end{tabular}
	}
	\vspace{-1cm}
	
	{\scriptsize
		{\setlength{\tabcolsep}{0.55em}
			\begin{tabular}{lp{1.2cm}p{0.1mm}p{0.0mm}p{0.0mm}p{0.00mm}p{0.00mm}p{0.00mm}p{0.00mm}p{0.00mm}p{0.00mm}p{0.00mm}p{0.00mm}p{0.00mm}p{0.00mm}p{0.00mm}p{0.00mm}p{0.00mm}p{0.00mm}p{0.00mm}}
				\multicolumn{2}{l}{$<$st$>$}	& 1  & \multicolumn{16}{l}{\multirow{12}{*}{\includegraphics[width=4.3cm,height=3.9cm]{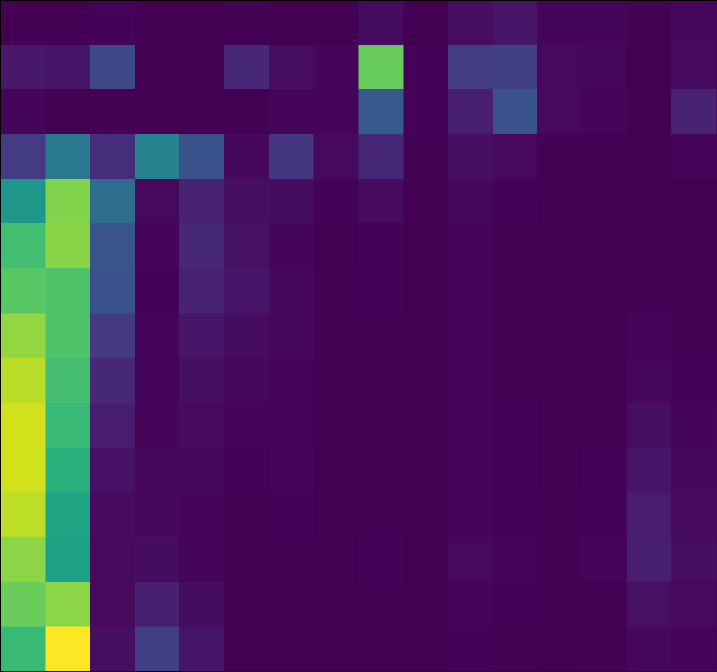}}}          \\[0.7pt]
				\multicolumn{2}{l}{return}	& 2  & \multicolumn{16}{l}{}                            \\[0.7pt]
				\multirow{9}{*}[17pt]{\Bigg\downarrow} & \multirow{10}{*}[23pt]{\makecell{\hspace{-0.3cm}\emph{predicting} \\ \hspace{-0.3cm}\emph{next word}}}	& 3  & \multicolumn{16}{l}{}                            \\[0.7pt]
				& & 4  & \multicolumn{16}{l}{}                            \\[0.7pt]
				& & 5  & \multicolumn{16}{l}{}                            \\[0.7pt]
				& & 6  & \multicolumn{16}{l}{}                            \\[0.7pt]
				& & 7  & \multicolumn{16}{l}{}                            \\[0.7pt]
				& & 8  & \multicolumn{16}{l}{}                            \\[0.7pt]
				&   & 9  & \multicolumn{16}{l}{}                            \\[0.7pt]
				&	& 10 & \multicolumn{12}{l}{}                            \\[0.7pt]
				&	& 11 & \multicolumn{12}{l}{}                            \\[0.7pt]
				&	& 12 & \multicolumn{12}{l}{}                            \\[0.7pt]
				&	& 13 & \multicolumn{12}{l}{}                            \\[0.7pt]
				&	&    & 1 & 2 & 3 & 4 & 5 & 6 & 7 & 8 & 9 & 10 & 11 & 12 & 13 & 14 & 15 & 16
			\end{tabular}
		}		
		
		\vspace{0.4cm}
		{\setlength{\tabcolsep}{0.55em}
			\begin{tabular}{lp{1.2cm}p{0.1mm}p{0.0mm}p{0.0mm}p{0.00mm}p{0.00mm}p{0.00mm}p{0.00mm}p{0.00mm}p{0.00mm}p{0.00mm}p{0.00mm}p{0.00mm}p{0.00mm}p{0.00mm}p{0.00mm}p{0.00mm}p{0.00mm}p{0.00mm}}
				\multicolumn{2}{l}{$<$st$>$}	& 1  & \multicolumn{16}{l}{\multirow{12}{*}{\includegraphics[width=4.3cm,height=3.9cm]{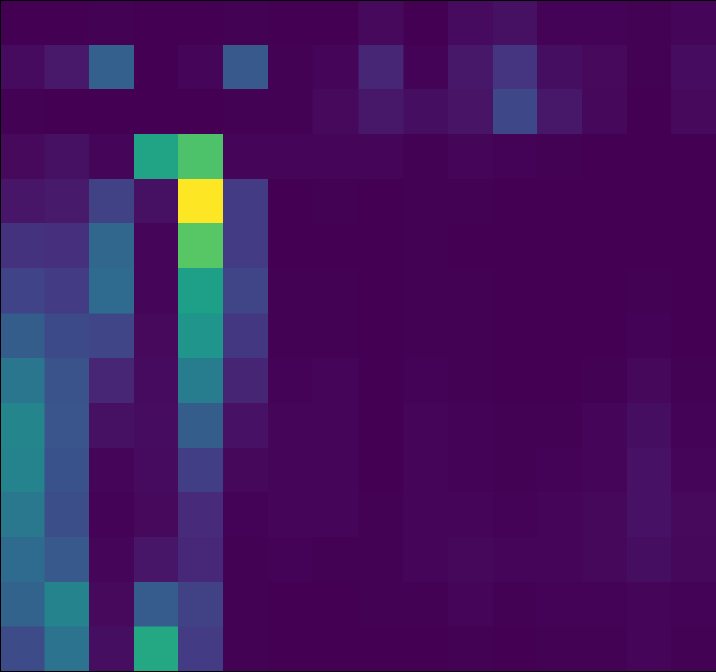}}}          \\[0.7pt]
				\multicolumn{2}{l}{lookup}	& 2  & \multicolumn{16}{l}{}                            \\[0.7pt]
				\multirow{9}{*}[17pt]{\Bigg\downarrow} & \multirow{10}{*}[23pt]{\makecell{\hspace{-0.3cm}\emph{predicting} \\ \hspace{-0.3cm}\emph{next word}}}	& 3  & \multicolumn{16}{l}{}                            \\[0.7pt]
				& & 4  & \multicolumn{16}{l}{}                            \\[0.7pt]
				& & 5  & \multicolumn{16}{l}{}                            \\[0.7pt]
				& & 6  & \multicolumn{16}{l}{}                            \\[0.7pt]
				& & 7  & \multicolumn{16}{l}{}                            \\[0.7pt]
				& & 8  & \multicolumn{16}{l}{}                            \\[0.7pt]
				&   & 9  & \multicolumn{16}{l}{}                            \\[0.7pt]
				&	& 10 & \multicolumn{12}{l}{}                            \\[0.7pt]
				&	& 11 & \multicolumn{12}{l}{}                            \\[0.7pt]
				&	& 12 & \multicolumn{12}{l}{}                            \\[0.7pt]
				&	& 13 & \multicolumn{12}{l}{}                            \\[0.7pt]
				&	&    & 1 & 2 & 3 & 4 & 5 & 6 & 7 & 8 & 9 & 10 & 11 & 12 & 13 & 14 & 15 & 16
			\end{tabular}
	}		}
	
	\vspace{0.1cm}
	\caption{Comparison of attention networks after predicting the action word in a baseline published at ICSE'19~\cite{leclair2019neural}.  The lower heatmap shows attention correctly applied after the correct action word is used.  The upper heatmap shows how the model attends to a different area of code, leading the model to produce incorrect prediction after an incorrect action word.}
	\label{fig:ex1}
\end{figure}

\section{Discussion and Conclusions}
\label{sec:conclusion}

In this paper, we introduce ``action word prediction'' as a complementary problem to source code summarization.  We argue that action word prediction is a key component of code summarization, because 1) high quality summaries tend to use an action word as the first word in the summary, and 2) the first word of a prediction tends to have a high impact on the subsequent predictions from a model.  A key point is that source code summarization tools very often need to predict the action word anyway, so a special emphasis on this problem is justified due to that word's importance.  We articulate our problem definition and supporting evidence in Section~\ref{sec:awp}.

During our experiments, we observed many cases in which incorrect action word prediction led to incorrect source code summary generation.  Consider Figure~\ref{fig:ex1} for one example.  The first 16 tokens in the preprocessed context (and the associated unpreprocessed code, for comparison) is visible.  Each position in the context is associated with a position in the x-axis of the heatmaps.  Each position in the summary is associated with a position in the y-axis.
The {\small \texttt{ast-attendgru}} baseline from ICSE'19~\cite{leclair2019neural} predicts the action word ``return.''  The upper heatmap is the attention network when predicting the next word: strong attention is visible in the first and second position, which is the return type (a char pointer).  The model predicts ``a pointer to'' as a result.  On the other hand, if we feed the model the correct action word (lookup), then the attention for the next word is on the fifth position, which is the start of the parameter list.  This attention makes sense because a ``lookup'' function is likely to look up something that has been passed in as a parameter.

Because of the impact of action word prediction on code summarization, we also argue that source code summarization tools should be evaluated in terms of action word prediction in addition to the overall prediction accuracy.  The traditional techniques for evaluation include automated scores such as BLEU and/or human studies.  The automated scores are calculations of overlap to a reference summary and, while helpful in giving an overall view of performance, do not provide a detailed view of where problems in a prediction occur.  For example, using the word ``an'' instead of ``a'' in a summary is considered BLEU to be an error of equal weight to using ``add'' and ``remove.''  On the other hand, human studies give a detailed view of errors, but are extremely expensive and are not feasible for evaluating small changes to a model or in providing quick feedback to a researcher.  For example, a grid search during parameter optimization may require dozens or hundreds of different configurations of the same model.  It is not practical to expect a human study for each of these configurations.  At the same time, using BLEU or similar scores may not provide a detailed enough picture to select the optimal configuration.

What we propose is that researchers evaluate the action word prediction quality of their source code summarization techniques in addition to, or even in certain cases in lieu of, the overall summary prediction quality.  We demonstrate in an experiment how this evaluation can be performed using common, well-understood metrics such as precision and recall.  We show this evaluation over datasets in different languages (C/C++ and Java) and conditions (standard and challenge).

For some cases such as challenge conditions, the problem of source code summarization may be too difficult to solve directly.  The situation is analogous to early AI problems in which chess playing algorithms were used as a wedge against larger, even more difficult problems~\cite{shannon1950xxii}.  A more recent example is in computer vision, how image classification started with simple examples of classifying very different objects e.g. faces versus furniture, before moving to more difficult problems~\cite{rawat2017deep}.  What we recommend is to use action word prediction as the ``wedge'' towards source code summarization.  The idea is that in very difficult conditions, a reasonable target problem is to correctly predict only the action word of a summary.  If the problem of action word prediction can be solved with high accuracy, then the research community would be better placed to solve code summarization more generally.

\section{Reproducibility}
\label{sec:appendix}

To encourage reproducibility and aid other research groups, we have released all data, source code, scripts, and tutorial information in an online appendix:

{\small \texttt{https://github.com/actionwords/actionwords}}

\section*{Acknowledgment}

This work is supported in part by NSF CCF-1452959 and CCF-1717607. Any opinions, findings, and conclusions expressed herein are the authors and do not necessarily reflect those of the sponsors.

\bibliographystyle{IEEEtran}
\bibliography{biblio}

\end{document}